\documentstyle[preprint,pra,aps,epsf]{revtex}

\begin{document}
\draft
\preprint{CfA 4436}
\title{Retarded long-range potentials
for the alkali-metal atoms and a perfectly conducting wall}
\author{M. Marinescu\protect\cite{MMpresentaddr}, A. Dalgarno, and J. F. Babb}
\address{
Harvard-Smithsonian Center for Astrophysics,\\
60 Garden Street, Cambridge, MA 02138
}
\date{Nov. 4, 1996}

\maketitle
\begin{abstract}
The retarded long-range potentials for hydrogen and alkali-metal atoms
in their ground states and a perfectly conducting wall are calculated.
The potentials are given over a wide range of atom-wall distances 
and the validity of the approximations used is established.
\end{abstract}
\narrowtext

The long-range interaction potential of an atom and a surface
characterizes the physics of many actual systems.  For ground state
alkali-metal atoms it has been measured by deflection of atomic
beams in scattering~\cite{ShiPar75}, reflection of atoms in atomic
fountains~\cite{KasMolRii91,LanCouLab96}, and by  combinations of
deflection and spectroscopy~\cite{Hin92,SukBosCho93}.  As the
atom-wall separation increases the interaction potential weakens due
to retardation arising from the finite speed of
light~\cite{Physics-Today,Mil93}.  A retarded potential has been
observed for Na atoms between parallel metallic
plates~\cite{SukBosCho93} and for the hydrogen atom inclusion of
retardation in theoretical calculations of hydrogen-liquid helium film
sticking coefficients dramatically improves the agreement with
experiment~\cite{YuDoySan93,CarCol92}.  In the present paper we
present calculations of the retarded long-range potentials for
hydrogen and alkali-metal atoms and a perfectly conducting wall.  When
a metallic wall has high reflectivity at the  atomic resonance 
transition wavelengths in the atomic electric dipole polarizability it
can be treated as a perfectly conducting wall~\cite{HinSan91} for
atom-wall separations large enough that higher multipoles in the
interaction and overlap with the surface electrons are unimportant.
Approximations proportional to the
the retarded  potential for an atom and a perfectly conducting  wall
have been given for the corresponding retarded potential
for a dielectric wall~\cite{DzyLifPit60,TikSpr93b}.
Thus the present results may be useful in analyzing the
interactions of atoms with dielectric or metallic walls for
spectroscopy or atom interferometry experiments~\cite{EksSchCha95}.

The retarded long-range potential $V(R)$ of a spherically-symmetric
ground state atom and a perfectly conducting wall, where $R$ is the
atom-wall distance, was derived in 1948 by Casimir and
Polder~\cite{CasPol48}.  It can be written~\cite{TikSpr93b}
\begin{equation}
\label{pot}
V(R) =  \frac{-1}{4 \pi\alpha_{\rm fs} R^4}
         \int_0^\infty dx\;\alpha (ix/\alpha_{\rm fs} R)  \exp(-2x)
         [2x^2+2x+1] ,
\end{equation}
where $\alpha (i\omega)$ is the atomic dynamic electric dipole
polarizability evaluated at imaginary frequency $i\omega$ and
$\alpha_{\rm fs}=1/137.035\,989\,5$ is the fine structure constant.
We use atomic units throughout.  
For small distances Eq.~(\ref{pot}) has the limiting form
\begin{equation}
\label{small-R}
V (R) \sim -C_3/R^3, \quad  R\rightarrow 0 ,
\end{equation}
where 
\begin{equation}
\label{C3-integral}
C_3 = \frac{1}{4\pi} \int_0^\infty d\omega  \alpha (i\omega) .
\end{equation}
As the distance increases,
retardation arising from the finite speed of light becomes
increasingly important and $V(R)$ approaches its asymptotic form, a
simple power law,
\begin{equation}
\label{asymp}
V(R) \sim V_\infty (R) \equiv -K_4/R^4 , \quad R\rightarrow\infty ,
\end{equation}
where
\begin{equation}
\label{casimir}
K_4 =  3 \alpha (0)/  8\pi\alpha_{\rm fs} = 16.36\, \alpha (0)  .
\end{equation}

A method has been developed to calculate directly the dynamic electric
dipole polarizability corresponding to an electron in the presence of
a central core field~\cite{MarSadDal94}.  It gives the function
$\alpha(i\omega)$ for hydrogen exactly.  Combined with a correction
reflecting the polarization of the alkali-metal core by the valence
electron~\cite{MarSadDal94,MarSadDal94b}, we used it with a model
potential for the field of the positive-ion core seen by the atomic
valence electron~\cite{MarSadDal94} to obtain the functions
$\alpha'(i\omega)$ for the alkali-metal atoms, where the prime has
been introduced to indicate the results for the alkali-metal atoms
obtained using this method. In determining the correction, measured
values of $\alpha (0)$ were adopted.  For Li, K, Rb, and Cs the values
of $\alpha (0)$ are those given in Ref.~\cite{MolSchMil74}.  For Na
the value $\alpha(0)=162.7$~\cite{EksSchCha95} was used, yielding a
value of the cutoff radius $r_c= 2.860\,812\,8$ in the core-polarization
correction compared to the values $\alpha (0)=159.2$ and
$r_c=0.379\,266\,0$ of Ref.~\cite{MarSadDal94}.  The values of $\alpha (0)$,
and the corresponding values of $K_4$, are listed 
in Table~\ref{C3-table}.

\begin{figure}[p]
\epsfxsize=1.\textwidth \epsfbox{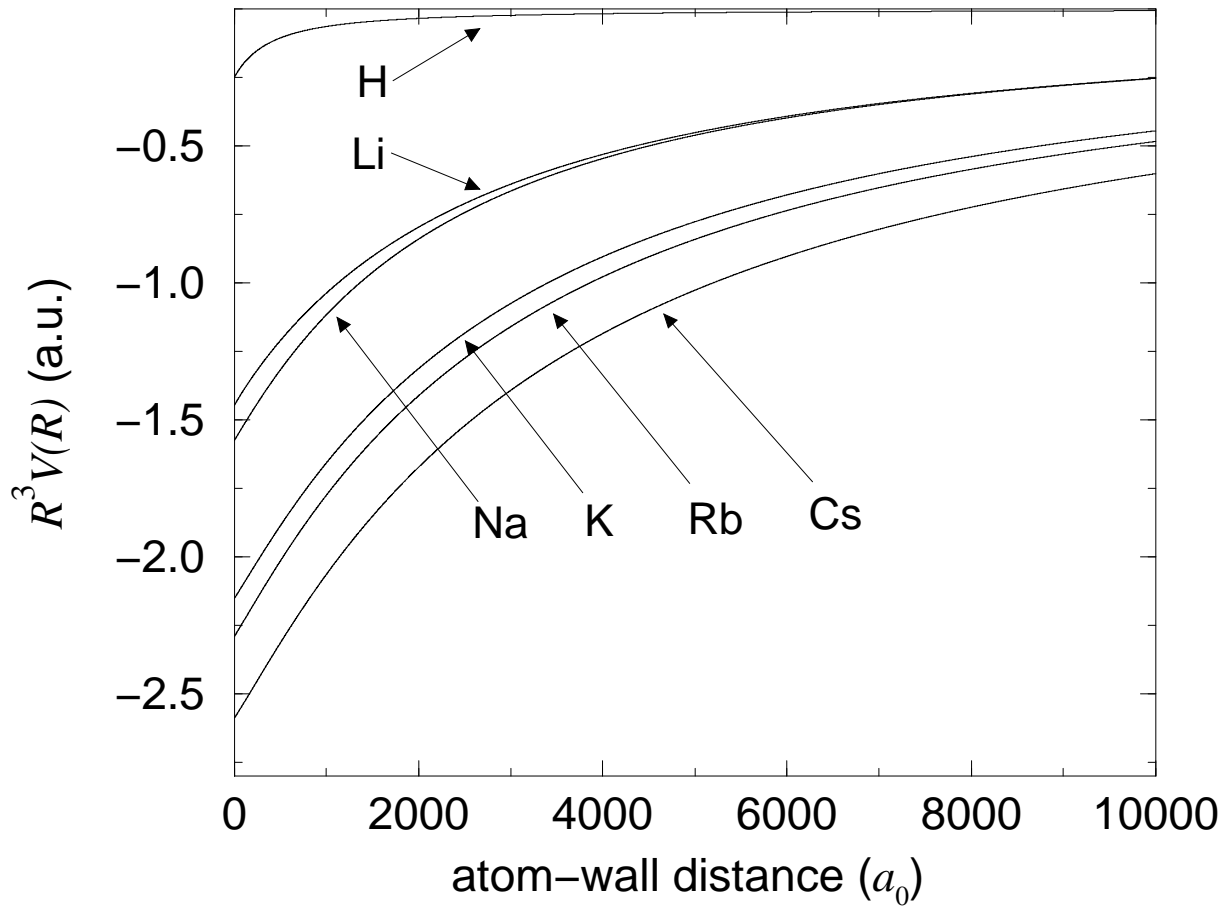}
\caption{For H, Li, Na, K, Rb, and Cs, 
values of $R^3 V(R)$, where $V(R)$ is the atom-wall potential,
Eq.~(\protect\ref{pot}), as a function of the atom-wall distance $R$ in
$a_0$.\label{alkali-fig}}
\end{figure}

The calculated values for $V (R)$ for hydrogen and the alkali-metal
atoms are given in Table~\ref{pot-table}.  
Eq.~(\ref{pot}) was integrated with a Gauss-Laguerre quadrature
method. The results for H
are exact.  In Fig.~\ref{alkali-fig} plots of $V(R)$ are presented for
hydrogen and the alkali-metal atoms.

Hinds and Sandoghdar~\cite{HinSan91} calculated the discrete valence
electron contributions to $\alpha (\omega)$ and evaluated the retarded
long-range energy shift for ground state Rb.  They presented the
results graphically for atom-wall distances from about 0.1 to
0.7$~\mu$m.  For example, they give a potential energy of $-1.15$~MHz
at 0.106$~\mu$m $(2000a_0)$, in agreement with our value of
$-1.162$~MHz, obtained from Table~\ref{C3-table} and
Figure~\ref{alkali-fig}, where the energy, in MHz, was obtained by
multiplying the value of $V(R)$ in a.u. by $6.580 \times 10^9 $.

The function $\alpha'(i\omega)$ is an  excellent approximation to
the exact dynamic polarizability 
$\alpha(i\omega)$  for photon energies $\omega$ greater than the first
excitation threshold of the core~\cite{Nor73,MaeKut79},
which means, accordingly, that $V(R)$ will be most accurate for values
of $R \gtrsim 1/(\alpha_{\rm fs}E_{\rm opt})$, where in atomic units,
$E_{\rm opt}$ is the  energy of the first excitation transition.
The model potential method does not completely account for
inner shell contributions to the oscillator strength distribution
and the resulting error in $\alpha'(i\omega)$ leads to 
a corresponding error in the evaluation of $V(R)$ for small $R$.

\begin{figure}[p]
\epsfxsize=1.\textwidth \epsfbox{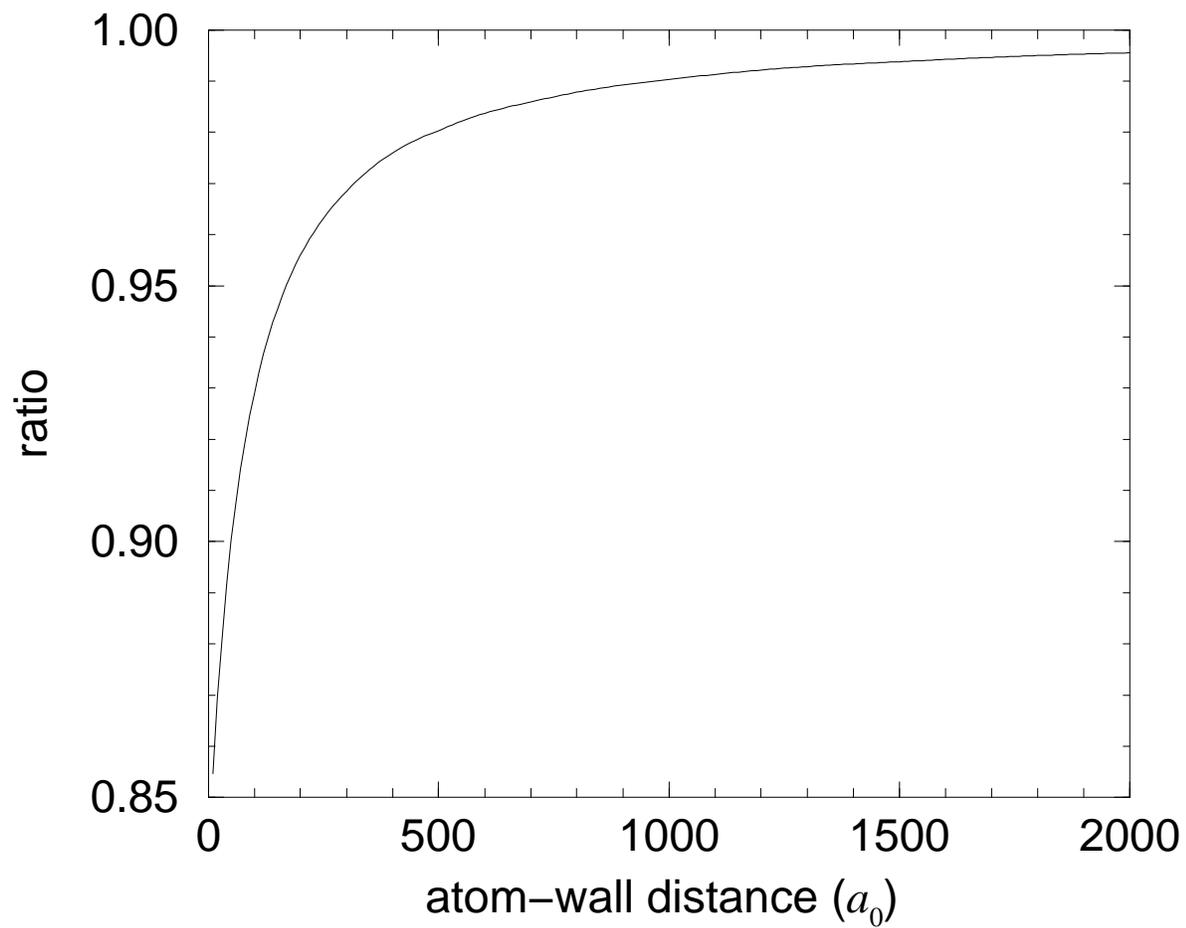}
\caption{For Na, ratio of
the atom-wall potential calculated in the present work to the more
accurate potential calculated by Kharchenko {\em et
al.\/}~\protect\cite{KhaBabDal96}.\label{fig-ratio}}
\end{figure}

For Na, Kharchenko {\em et al.\/}~\cite{KhaBabDal96} constructed a
semi-empirical representation of the function $\alpha(i\omega)$, which
explicitly included core contributions, using a combination of
theoretical and experimental photoionization and photoabsorption cross
sections, and evaluated $V(R)$ using Eq.~(\ref{pot}).  To determine
the accuracy of the potentials in the present work, the ratio of the
potential obtained using $\alpha'(i\omega)$ to the potential obtained
using the empirical $\alpha(i\omega)$ was calculated. It is given in
Fig.~\ref{fig-ratio}.  For Na, the optical transition energy is
$E_{\rm opt} \sim 0.077$, suggesting that the ratio should be near
unity for $R\gtrsim 1800$, which it is, as seen in
Fig.~\ref{fig-ratio}.  The potential calculated using
$\alpha'(i\omega)$ underestimates the true potential by 5\% at $R=200$
and by 2\% at $R=500$.

J.F.B. would like to thank L. Spruch, E.~Hinds, D.~Pritchard, and
T. Hammond for useful discussions.  This work was supported in part
(M.M. and A.D.) by the U.S. Department of Energy, Division of Chemical
Sciences, Office of Basic Energy Sciences, Office of Energy Research
and the Smithsonian Institution.  The Institute for Theoretical Atomic
and Molecular Physics is supported by a grant from the National
Science Foundation to the Smithsonian Institution and Harvard
University.
\begin{table}
\begin{center}
\caption{
Static electric dipole polarizabilities $\alpha (0)$
for the atoms and asymptotic Casimir coefficients $K_4$ for the
alkali-atom-wall interactions.}
\label{C3-table}
\begin{tabular}{llllllll}
\multicolumn{1}{c}{Quantity} &\multicolumn{1}{c}{Ref.} &
   \multicolumn{1}{c}{H}  & \multicolumn{1}{c}{Li} &
   \multicolumn{1}{c}{Na} & \multicolumn{1}{c}{K}  & 
   \multicolumn{1}{c}{Rb} &  \multicolumn{1}{c}{Cs} \\
\hline
$\alpha (0)/a_0^3$ 
      & Present~\tablenote{Same as Ref.~\protect\cite{MarSadDal94}, 
  Table~II, line labeled ``b'', except for Na.}
  & 4.500 & 164.0 & 162.7 & 292.8 & 319.2 & 402.2  \\
$K_4/a_0^4$ 
      & Present & 73.62  & 2683  & 2662  & 4789 & 5221 & 6579 \\
\end{tabular}
\end{center}
\end{table}
\begin{table}
\begin{center}
\caption{
Values of $-R^3 V(R)$, where $V(R)$
is the atom-wall potential, Eq.~(\protect\ref{pot}),
for the hydrogen-wall and alkali-atom-wall interactions, in atomic
units. Numbers in square brackets represent powers of ten.}
\label{pot-table}
\begin{tabular}{lllllll}
\multicolumn{1}{c}{$R/a_0$} & \multicolumn{1}{c}{H} & \multicolumn{1}{c}{Li} &
   \multicolumn{1}{c}{Na} & \multicolumn{1}{c}{K} & \multicolumn{1}{c}{Rb} &
   \multicolumn{1}{c}{Cs} \\
\hline
1.0    &  2.4981[-1] &  1.4468     &  1.5753     &  2.1521     &  2.2903     &  2.5882     \\
1.5    &  2.4951[-1] &  1.4466     &  1.5750     &  2.1518     &  2.2900     &  2.5879     \\
2.0    &  2.4912[-1] &  1.4462     &  1.5746     &  2.1514     &  2.2896     &  2.5875     \\
2.5    &  2.4868[-1] &  1.4457     &  1.5742     &  2.1510     &  2.2892     &  2.5871     \\
3.0    &  2.4824[-1] &  1.4453     &  1.5737     &  2.1506     &  2.2888     &  2.5867     \\
4.0    &  2.4742[-1] &  1.4445     &  1.5729     &  2.1498     &  2.2880     &  2.5860     \\
5.0    &  2.4674[-1] &  1.4438     &  1.5723     &  2.1492     &  2.2873     &  2.5854     \\
6.0    &  2.4616[-1] &  1.4432     &  1.5717     &  2.1486     &  2.2868     &  2.5849     \\
8.0    &  2.4518[-1] &  1.4422     &  1.5707     &  2.1477     &  2.2859     &  2.5840     \\
1.0[1] &  2.4423[-1] &  1.4412     &  1.5698     &  2.1468     &  2.2849     &  2.5831     \\
1.5[1] &  2.4158[-1] &  1.4385     &  1.5672     &  2.1443     &  2.2824     &  2.5806     \\
2.0[1] &  2.3872[-1] &  1.4356     &  1.5643     &  2.1415     &  2.2795     &  2.5779     \\
2.5[1] &  2.3589[-1] &  1.4326     &  1.5614     &  2.1387     &  2.2767     &  2.5752     \\
3.0[1] &  2.3316[-1] &  1.4298     &  1.5586     &  2.1360     &  2.2739     &  2.5725     \\
4.0[1] &  2.2795[-1] &  1.4242     &  1.5530     &  2.1306     &  2.2684     &  2.5671     \\
5.0[1] &  2.2295[-1] &  1.4187     &  1.5475     &  2.1252     &  2.2629     &  2.5618     \\
6.0[1] &  2.1811[-1] &  1.4132     &  1.5420     &  2.1198     &  2.2573     &  2.5564     \\
8.0[1] &  2.0892[-1] &  1.4024     &  1.5309     &  2.1089     &  2.2462     &  2.5456     \\
1.0[2] &  2.0038[-1] &  1.3919     &  1.5199     &  2.0981     &  2.2349     &  2.5346     \\
1.5[2] &  1.8156[-1] &  1.3662     &  1.4927     &  2.0708     &  2.2068     &  2.5070     \\
2.0[2] &  1.6567[-1] &  1.3417     &  1.4659     &  2.0436     &  2.1787     &  2.4791     \\
2.5[2] &  1.5211[-1] &  1.3180     &  1.4397     &  2.0166     &  2.1507     &  2.4512     \\
3.0[2] &  1.4042[-1] &  1.2951     &  1.4141     &  1.9898     &  2.1228     &  2.4233     \\
4.0[2] &  1.2136[-1] &  1.2515     &  1.3646     &  1.9372     &  2.0681     &  2.3680     \\
5.0[2] &  1.0653[-1] &  1.2106     &  1.3176     &  1.8862     &  2.0148     &  2.3136     \\
6.0[2] &  9.4707[-2] &  1.1720     &  1.2729     &  1.8368     &  1.9632     &  2.2606     \\
8.0[2] &  7.7144[-2] &  1.1011     &  1.1905     &  1.7433     &  1.8654     &  2.1590     \\
1.0[3] &  6.4810[-2] &  1.0374     &  1.1165     &  1.6568     &  1.7747     &  2.0636     \\
1.5[3] &  4.5911[-2] &  9.0333[-1] &  9.6172[-1] &  1.4683     &  1.5762     &  1.8515     \\
2.0[3] &  3.5345[-2] &  7.9700[-1] &  8.4053[-1] &  1.3130     &  1.4122     &  1.6728     \\
2.5[3] &  2.8659[-2] &  7.1098[-1] &  7.4384[-1] &  1.1839     &  1.2754     &  1.5214     \\
3.0[3] &  2.4069[-2] &  6.4023[-1] &  6.6535[-1] &  1.0755     &  1.1601     &  1.3923     \\
4.0[3] &  1.8199[-2] &  5.3139[-1] &  5.4651[-1] &  9.0469[-1] &  9.7795[-1] &  1.1849     \\
5.0[3] &  1.4616[-2] &  4.5218[-1] &  4.6156[-1] &  7.7725[-1] &  8.4153[-1] &  1.0269     \\
6.0[3] &  1.2206[-2] &  3.9236[-1] &  3.9829[-1] &  6.7922[-1] &  7.3628[-1] &  9.0342[-1] \\
8.0[3] &  9.1748[-3] &  3.0869[-1] &  3.1107[-1] &  5.3949[-1] &  5.8578[-1] &  7.2426[-1] \\
1.0[4] &  7.3473[-3] &  2.5346[-1] &  2.5428[-1] &  4.4556[-1] &  4.8430[-1] &  6.0173[-1] \\
1.5[4] &  4.9032[-3] &  1.7400[-1] &  1.7361[-1] &  3.0815[-1] &  3.3541[-1] &  4.1944[-1] \\
2.0[4] &  3.6787[-3] &  1.3200[-1] &  1.3140[-1] &  2.3452[-1] &  2.5542[-1] &  3.2034[-1] \\
2.5[4] &  2.9435[-3] &  1.0619[-1] &  1.0559[-1] &  1.8897[-1] &  2.0587[-1] &  2.5859[-1] \\
3.0[4] &  2.4531[-3] &  8.8763[-2] &  8.8202[-2] &  1.5811[-1] &  1.7228[-1] &  2.1659[-1] \\
4.0[4] &  1.8400[-3] &  6.6780[-2] &  6.6313[-2] &  1.1906[-1] &  1.2976[-1] &  1.6329[-1] \\
5.0[4] &  1.4721[-3] &  5.3501[-2] &  5.3110[-2] &  9.5433[-2] &  1.0402[-1] &  1.3095[-1] \\
6.0[4] &  1.2267[-3] &  4.4620[-2] &  4.4286[-2] &  7.9611[-2] &  8.6777[-2] &  1.0927[-1] \\
8.0[4] &  9.2008[-4] &  3.3492[-2] &  3.3236[-2] &  5.9772[-2] &  6.5156[-2] &  8.2066[-2] \\
9.0[4] &  8.1785[-4] &  2.9777[-2] &  2.9548[-2] &  5.3146[-2] &  5.7934[-2] &  7.2975[-2] \\
1.0[5] &  7.3607[-4] &  2.6804[-2] &  2.6597[-2] &  4.7842[-2] &  5.2153[-2] &  6.5696[-2] \\
1.0[6] &  7.3609[-5] &  2.6826[-3] &  2.6613[-3] &  4.7894[-3] &  5.2212[-3] &  6.5788[-3] \\
\end{tabular}
\end{center}
\end{table}

\clearpage
\clearpage

\clearpage


\end{document}